# To Seal or Not To Seal

Javad Eshraghi[1], Sunghwan Jung[2], Pavlos P. Vlachos[1,*]

[1]Department of Mechanical Engineering, Purdue University, West Lafayette, IN 47907
[2]Department of Biological and Environmental Engineering, Cornell University, Ithaca, NY 14853
*To whom correspondence should be addressed. E-mail: pvlachos@purdue.edu

When an object impacts the free surface of a liquid, it ejects a splash curtain upwards and creates an air cavity below the free surface. As the object descends into the liquid, the air cavity eventually closes under the action of hydrostatic pressure (deep seal). In contrast, the surface curtain may splash outwards or dome over and close, creating a surface seal. In this paper we experimentally investigate how the splash curtain dynamics are governed by the interplay of cavity pressure difference, gravity, and surface tension and how they control the occurrence, or not, of surface seal. Based on the experimental observations and measurements, we develop an analytical model to describe the trajectory and dynamics of the splash curtain. The model enables us to reveal the scaling relationship for the dimensionless surface seal time and discover the existence of a critical dimensionless number that predicts the occurrence of surface seal. This scaling indicates that the most significant parameter governing the occurrence of surface seal is the velocity of the airflow rushing into the cavity. This is in contrast to the current understanding which considers the impact velocity as the determinant parameter.

## I. Introduction

Work on water entry has been predominatly focused on the dynamics and fluid motion that occur below the free surface [1-6], often not considering the role of surface seal [1] or limiting the work to cases without it [7]. Although surface seal and deep seal are driven by different physical mechanisms and result in different cavity dynamics, the phenomena occurring above the free surface are inherently linked to those occurring below the free surface [8, 9]. In this paper, we focus on the splash curtain to investigate the factors that govern a surface seal.

To study the dynamics of the splash curtain we drop different spheres into a 25×40×50 cm$^3$ glass-sided aquarium tank filled with distilled water and record the phenomena using a high-speed digital video camera operating at 5000 fps. The tank does not interfere with cavity expansion or pinch-off for any of the experiments reported. The dropping mechanism is mounted above the tank to release spheres without imparting spin, which can strongly affect the cavity dynamics [5].

The projectiles used here include Acrylic ($\rho_s = 1.18\ g/cm^3$), glass ($\rho_s = 2.40\ g/cm^3$), alumina ($\rho_s = 3.96\ g/cm^3$), steel ($\rho_s = 7.87\ g/cm^3$), and tungsten ($\rho_s = 19.30\ g/cm^3$) spheres of different diameters (d = 9.525–19.05 mm). In order to produce a cavity, the projectiles are coated with WX2100 [8, 10], which creates a hydrophobic surface condition with a contact angle of 150°–165°. The sphere impact velocities ranged from 2.0 m/s to 6.0 m/s (by 0.5 m/s increments). The impact of a projectile of radius $R_0$ into a liquid at velocity $U_0$ is characterized by the non-dimensional Froude number $Fr = U_0/\sqrt{gR_0}$, where $g$ is the gravitational acceleration [11, 12]. The Reynolds number ($Re$), the Weber number ($We$), and the Bond number

($Bo$), are also often used to provide insight into the interplay and relative importance of the physical forces governing free surface interaction in the water entry phenomenon; $Re = R_0 U_0/\nu$ where $\nu$ is the kinematic viscosity of the liquid, $We = \rho U^2 R_0/\gamma$ where $\rho$ is the liquid density and $\gamma$ is the surface tension, $Bo = \Delta\rho g R_0^2/\gamma$ where $\Delta\rho$ is the density difference.

## II. Experimental observations

During the initial stages of impact, a splash curtain is ejected upwards and outwards, as seen in Fig. 1 [13]. In addition, as the sphere descends into the fluid, an expanding air cavity is formed behind it [14].

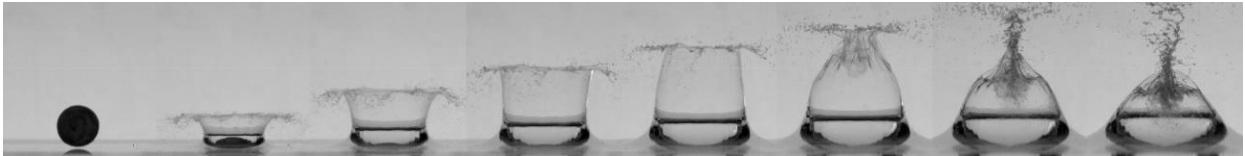

Figure 1: Splash curtain formed by a steel sphere, $R_0 = 0.95$ cm, $U_0 = 5.5$ m/s, $\Delta t = 3$ ms.

As the splash rises from the surface, it is subject to two main forces leading to its collapse: reduction in pressure caused by the airflow entrained into the cavity behind the sphere, and surface tension [15, 16]. The air flowing into the expanding cavity induces pressure drag acting on the splash curtain and draws the splash radially inward. Surface closure is important in the development of the cavity and influences the later cavity growth [9]. After surface closure, the cavity continues to expand due to the inertial effects of the sphere moving through the fluid and the pressure inside the cavity decreases. The pressure governing the deep seal in the impact-produced cavities is the sum of the hydrostatic pressure due to the depth [11, 16] and the pressure deficit in the cavity [15, 17]. The greater the pressure difference between cavity and air above the surface, the closer the pinch-off location will be to the surface [17, 18].

Our observations show that the cavity characteristics are highly dependent on the sphere density and do not scale linearly with $Fr$ as had been observed experimentally by [19] for low $Fr$; The deviation is most pronounced for the cases with surface seal. Even though the modified expressions for cavity characteristics developed by [7] agree with experimental observations in the regime without surface seal, they are not reliable in the surface seal regime, since the effect of surface seal on the cavity is ignored [20].

In the closure of splash curtain, an intuitive interpretation might label an impact velocity as the main cause in the occurrence of surface seal phenomenon. However, experimental observations imply dependency of surface seal not only on the sphere impact velocity, but also on the sphere size and density. To identify the surface closure mechanism, we will look at the interactions between the forces acting on the splash curtain, pulling it inward.

## III. Splash curtain modeling

Shortly after the splash is ejected upon impact, surface tension causes the fluid at the tip of the splash to coalesce, forming a rounded rim. This rim is approximated as an axisymmetric circular ring about the z-axis, attached to a thin fluid film. In reality, the splash is very irregular, often forming a crown-like splash similar to those observed by [21, 22].

The position of this rim is described by the vector $\vec{x}(t) = r(t)\hat{r} + z(t)\hat{z}$, in the $r$-$z$ plane. Figure 2(a) shows the splash geometry and coordinate system just after impact. Normal and tangential coordinates (n and s, respectively) are also defined relative to the rim, in the direction of its instantaneous velocity. The angle $\theta$ is defined as the angle from the $\hat{r}$ unit vector to the $\hat{s}$ unit vector. Figure 2(b) shows the splash curtain at some later time as the trajectory of the splash has evolved. As the rim's trajectory evolves, the $\hat{n}$ and $\hat{s}$ coordinates remain fixed to the rim and their orientation is described by the angle $\theta$.

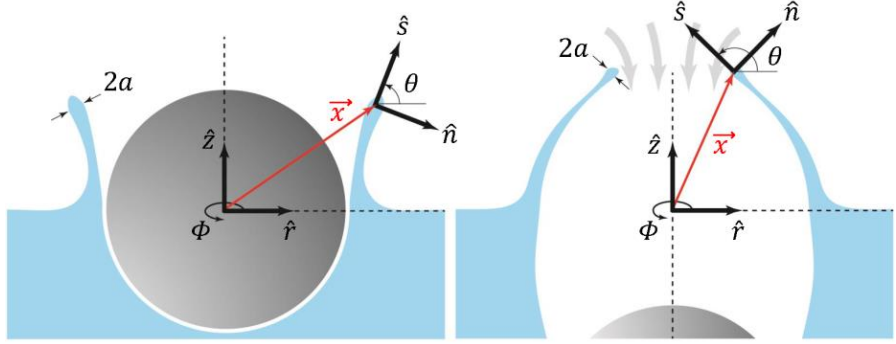

Figure 2: The coordinate system defined for the splash curtain trajectory model: (a) a time just after impact, (b) a later time just before surface seal.

The rounded rim is approximated by a finite mass, constant in time, with a circular profile in the $r$-$z$ plane. The rim radius, $a$, is obtained from the experiment and assumed to be constant in time (See Appendix 2). With the assumed geometry and the coordinate system shown in Fig. 2, a differential volume of the rim is given by $\rho \pi a^2 r d\phi$. This small mass is moving in a curved trajectory and has a centrifugal acceleration. This acceleration creates a force given by

$$\vec{F}_c(t) = \rho \pi a^2 r(t) d\phi \frac{|\dot{\vec{x}}(t)|^2}{R_c(t)} \hat{n} \tag{1}$$

where $R_c(t)$ is the instantaneous radius of curvature of the rim's trajectory given by

$$R_c(t) = \frac{d\theta(t)}{ds} \tag{2}$$

A drag force opposes the rim's motion, as the ejected fluid travels through the surrounding air. The drag force is given by

$$\vec{F}_d(t) = -\frac{1}{2}\rho_a C_d 2ar(t)d\phi|\dot{\vec{x}}(t)|^2 \hat{s} \tag{3}$$

where $C_d$ is the drag coefficient. There have been many attempts to model the drag coefficient on drops and thin liquid sheets [23]. For simplicity we have chosen to approximate the airflow around the rounded rim to be laminar and the drag coefficient may be expressed as $C_d = 24/Re$.

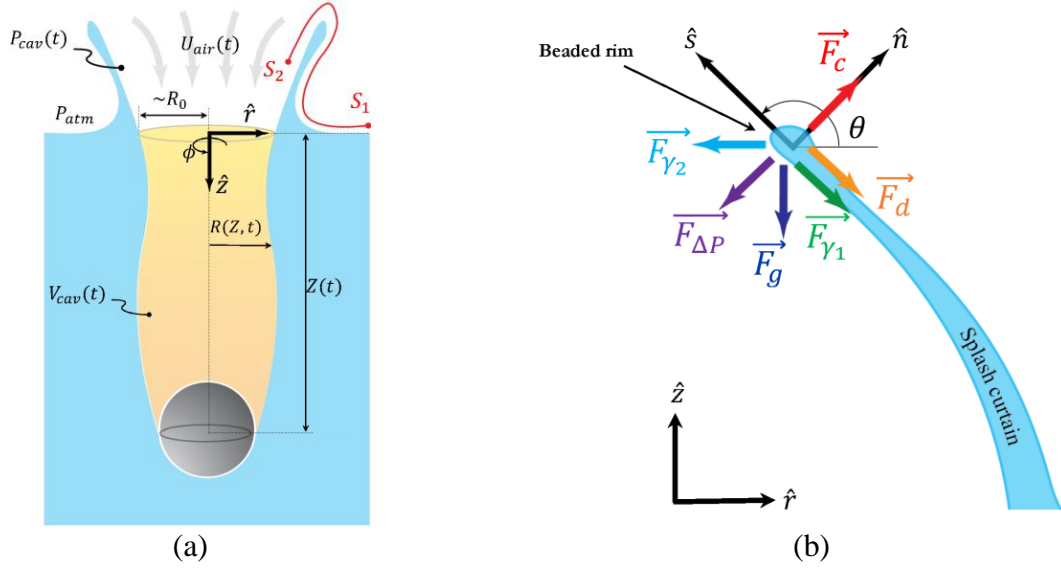

(a) (b)

Figure 3: (a) Cavity behind the sphere. The cavity volume, $V_{cav}(t)$, can be approximated by integrating the cavity profile, $R(z,t)$, over the length of the cavity, $Z(t)$. (b) Direction of the forces acting on the rounded rim of the splash curtain.

There are two surface tension forces that need to be considered [24, 25]. The first is a result of the thin fluid sheet attached to the rounded rim, given by

$$\vec{F}_{\gamma 1}(t) = 2\gamma r(t)d\phi \hat{s} \tag{4}$$

The sheet pulls the rim along the $s$ direction and we must account for both sides of the thin film (thus the factor of 2). The second surface tension force acts along the circumference of the rim, radially inward, given by

$$\vec{F}_{\gamma 2}(t) = -4\pi a \gamma d\phi \hat{r} \tag{5}$$

There is a gravitational force acting in the $z$ direction [24, 25], given by

$$\vec{F}_g(t) = -\rho g \pi a^2 r(t) d\phi \hat{z} \tag{6}$$

The expanding air cavity creates a pressure difference across the splash curtain. This pressure difference acts normal to the splash trajectory, collapsing it inward. This pressure difference creates a force given by

$$\vec{F}_{\Delta P}(t) = -2ar(t)d\phi \Delta P(t)\hat{n} \tag{7}$$

where $\Delta P(t)$ is pressure difference across the splash. This pressure difference can be estimated by measuring the time rate of change of cavity volume, $dV_{cav}(t)/dt$. If we know the area of the opening through which the air flows, a mean air velocity, $U_{air}(t)$, can be estimated (See Appendix 1), and from there the pressure drop can be estimated using Bernoulli's principle (Fig. 3(a)). This assumes the air flowing into the cavity is incompressible, viscous effects are negligible, and treats the cavity as a spatially uniform body. We will also assume that $P(t)$ is spatially uniform within the cavity. Figure 3(b) shows these forces and their directions relative to the splash curtain.

Summing these forces results in an equation of motion for the rim, given by

$$\rho \pi a^2 r(t) d\phi \frac{d^2 \vec{x}(t)}{dt^2} = \vec{F}_c + \vec{F}_{\gamma 1} + \vec{F}_{\gamma 2} + \vec{F}_g + \vec{F}_d + \vec{F}_{\Delta P} \tag{8}$$

Simplifying Eq. 8 results in a second order nonlinear ordinary differential equation describing the motion of the splash rim. The splash originates at a radial distance $R_0$ at the free surface ($z = 0$). The initial conditions in space are given by $\vec{x}(t=0) = R_0 \hat{r}$. The initial conditions for velocity are given by $\dot{\vec{x}}(t=0)$, which is obtained from the experiment.

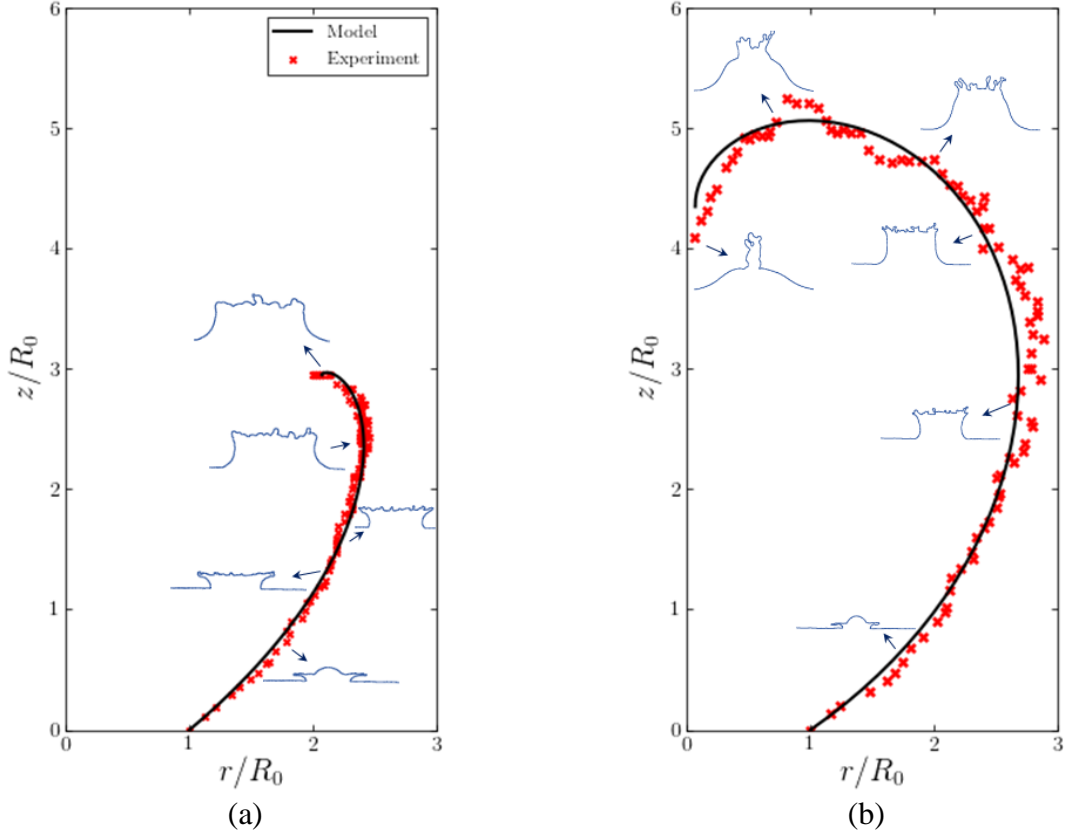

(a)          (b)

Figure 4: Splash curtain rounded rim trajectory: Experimental observation vs. Model prediction. (a) $We \approx 800$, $Bo \approx 12$, $DR = 7.87$, $U_{air}/U_0 \approx 0.1063$, (b) $We \approx 3950$, $Bo \approx 12$, $DR = 7.87$, $U_{air}/U_0 \approx 0.9474$ [26]. Density Ratio: $DR = \rho_s/\rho$, where $\rho$ is the fluid density, and $\rho_s$ is sphere density.

For a chosen set of system parameters and initial conditions, the trajectory of the splash curtain can be modeled by solving Eq. 8. To validate the proposed model, we compare the rim trajectory predicted by model with experimental observations in Fig. 4 [26, 27]. We define the first inflection point found in the splash curtain profile from the bottom of the curtain as the curtain rounded rim in the experimental videos (See Appendix 2).

Surface seal occurs when $r(t) = 0$. At this moment, the rounded rim has only one velocity component in the negative $\hat{z}$ direction. Thus, the occurrence of surface seal, is defined by two criteria given by

$$t_{surf} \rightarrow \dot{r}(t) = 0 \ \& \ \dot{z}(t) < 0 \tag{9}$$

where $t_{surf}$ is the time elapsed between the initial impact of the sphere and the surface seal. The model predicts the surface seal as when the two criteria in Eq. 9 are met. In the cases without surface seal, the time it takes for the splash to reach its maximum height is considered as the equivalent seal time and the same criteria (Eq. **9**) are applied for modeling of these cases as well.

## IV. Discussion

The model performance is most sensitive to the pressure force. In the establishment of the pressure difference, we used the airflow velocity. Another approach for the formulation would be replacing the airflow velocity with the sphere impact velocity. Figure 5 shows a comparison between the surface seal time predicted by the model and the measured value from the experiment as well as the model performance sensitivity to the velocity selection in pressure difference formulation. Figure 5(a) indicates that the surface seal time predicted by the constant pressure difference, defined based on the sphere impact velocity, does not agree with the experimental measurements. On the other hand, by taking into account the instantaneous pressure difference across the splash curtain (Fig. 5(b)) which is a representation of the cavity expansion history and airflow velocity, the maximum model error in the prediction of surface seal time is 5.6%.

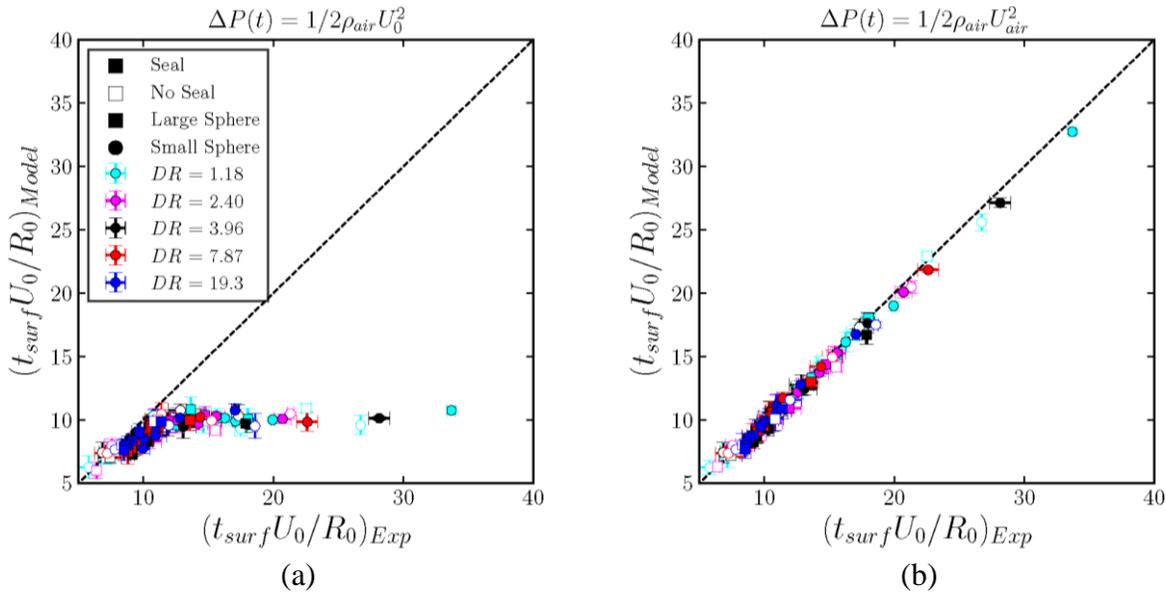

(a) (b)

Figure 5: Model prediction for dimensionless surface seal time vs. measured value for two different approaches in pressure difference estimation. In (a), a pressure difference was defined based on the sphere impact velocity which was constant during the impact and closure, while in (b), the pressure difference was estimated using the cavity volume method (See Appendix 1). Cases without surface seal are shown with open markers. In these cases, time it takes for the splash to reach its maximum height has been considered as the equivalent seal time.

In addition, we report the dimensionless surface seal time as a function of $We$, defined based on sphere impact velocity, in Fig. 6(a). In spite of showing the general decaying trend with $We$, this scaling fails to identify a critical $We$ for transition from no surface seal regime to surface seal regime for all the spheres with different density and size [28]. Other proposed scalings for the surface seal time also failed obtaining a threshold value for the occurrence of the surface seal [16-18].

These observations indicate that the sphere impact velocity is not the most significant parameter in the surface closure process. Based on the experiment, water entry cavities with quickly increasing volumes (high $dV_{cav}(t)/dt$ values) In other words, surface seal cavities typically are associated with surface seal. undergo a rapid increase in pressure difference (across the splash curtain) before dome closure, while a more gradual increase in pressure difference leads to a cavity without surface seal. Therefore, we can hypothesize that the closure dynamics of the splash is dominated by the pressure difference across it generated by cavity expansion and airflow into the cavity. Thus, to find the proper scaling for the dimensionless surface seal time, we only use pressure difference force to rewrite the equation of motion in the radial direction, $\hat{r}$, for the whole splash curtain

$$m \frac{d^2 \vec{R}(t)}{dt^2} = \frac{1}{2} \rho A U_{air}^2 \qquad (10)$$

where $m$ is the splash curtain mass and can be approximated as, $\rho A w$. Here, $A$ and $w$ are the splash curtain surface area and thickness, respectively. Substitution leads to

$$\frac{d^2 \vec{R}(t)}{dt^2} = \frac{1}{2w} U_{air}^2 \approx const. \qquad (11)$$

Solving Eq. **11** for $\vec{R}(t)$ yeilds

$$\vec{R}(t) = R_0 - \frac{1}{4w} U_{air}^2 t^2. \qquad (12)$$

At the instant of surface seal, $\vec{R}(t = t_{surf}) = 0$, and we find $t_{surf} = \sqrt{4wR_0}/U_{air}$. Assuming that the curtain thickness is a linear function of the sphere radius [29], $w \propto R_0$, surface seal time is deduced to be proportional to

$$t_{surf} \propto \frac{R_0}{U_{air}} \qquad (13)$$

This suggests that the proper scaling for dimensionless surface seal time is defined based on the velocity of airflow into the cavity

$$\frac{t_{surf} U_0}{R_0} \propto \frac{1}{\frac{U_{air}}{U_0}} \qquad (14)$$

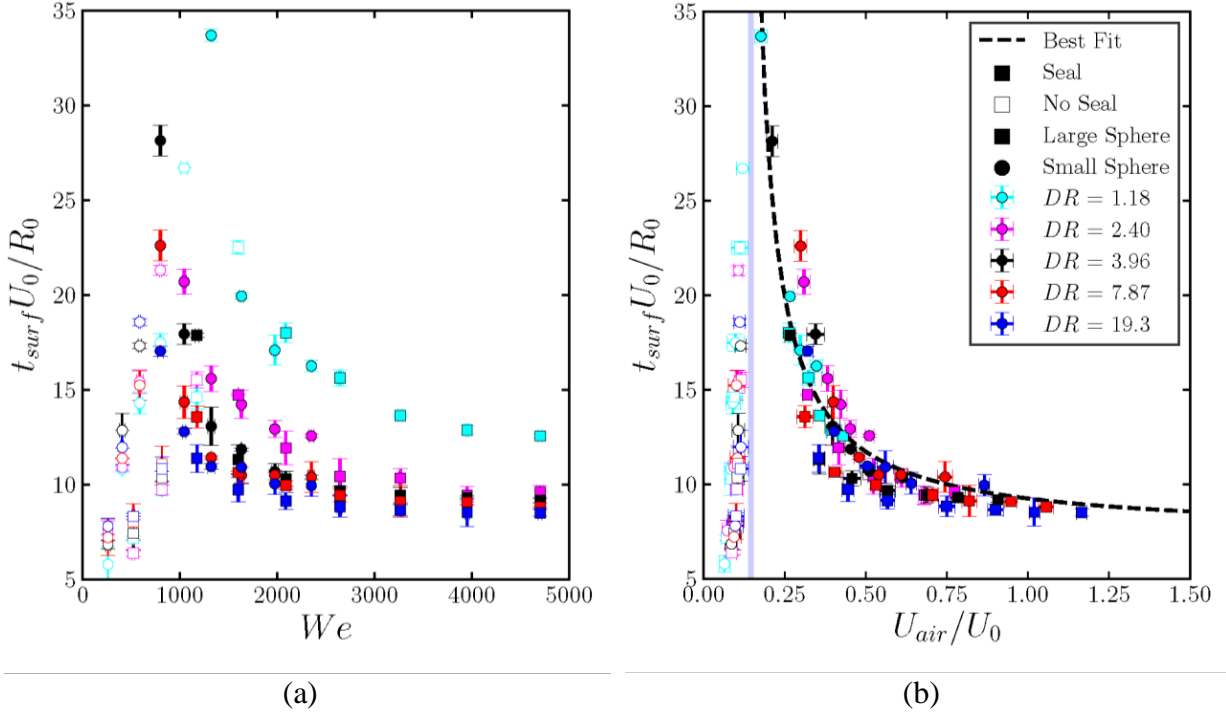

Figure 6: Dimensionless surface seal time, $t_{surf}U_0/R_0$, as a function of (a) $We$ defined based on sphere impact velocity, $We = \frac{\rho_{water}U_0^2 R_0}{\sigma}$, (b) ratio of air entrainment velocity to the sphere impact velocity. The dashed line is the fitted function in the form of Eq. **14**: $t_{surf}U_0/R_0 \approx \frac{1.68}{\frac{U_{air}}{U_0}-0.12} + 7.35$ with $R^2 \approx 0.89$. The predicted critical air entrainment velocity ratio from the mathematical model (Eq. **17**) is $U_{air}/U_0 \approx 0.146$ with uncertainty of $\approx \pm 0.005$ (calculated based on the rounded rim radius, $a$, uncertainty) as indicated by blue shaded area. Cases without surface seal are shown with open markers. In these cases, time it takes for the splash to reach its maximum height has been considered as the equivalent seal time.

Thus, we plot the dimensionless surface seal time as a function of air entrainment velocity to impact velocity ratio in Fig. 6(b). This plot indicates that this scaling enables us to obtain one single transition air velocity ratio from no surface seal regime to surface seal regime for all the spheres with different density and size. According to the fitted function equation (dashed line in Fig. 6(b)), the critical air velocity ratio for the occurrence of surface seal is 0.12.

Now, the model is used to find the critical airflow velocity beyond which the surface seal occurs. By revisiting Eq. **8** and conducting a scaling analysis, this equation is reduced to

$$\frac{d^2 R(t)}{dt^2} = -\frac{1}{\pi a}U_{air}^2 + \frac{2\gamma}{\rho \pi a^2}. \tag{15}$$

We solve Eq. **15** for $R(t)$ and set it equal to zero to find the surface seal time

$$t_{surf} = \sqrt{\frac{R_0}{\left[\frac{R_0}{\pi a U_0^2}U_{air}^2 - \frac{2\gamma R_0}{\rho \pi a^2 U_0^2}\right]}} \tag{16}$$

This equation has a real solution when the denominator is greater than zero and, in that case,

the occurrence of surface seal is assured. Therefore, the singularity of Eq. **16** corresponds to the transition from no seal regime to surface seal regime and it is formulated as

$$\frac{U_{air}^2}{U_0^2} = \frac{2\gamma}{\rho a U_0^2} = \frac{2}{We_c}\frac{R_0}{a} \qquad (17)$$

where $We_c$ is the critical Weber number for the occurrence of surface seal, defined based on sphere impact velocity. Experiments show that most of the cases with a $We$ greater than 1000 are associated with surface seal. Using experimental estimation for rounded rim radius, $a \approx 0.095 R_0$ ($a/R_0 \approx 0.095 \pm 0.007$), and choosing $We_c \approx 1000$, we can establish $U_{air}/U_0 \approx 0.146$ as a criteria for the occurrence of surface seal (Eq. **17**), which is consistent with the experiment (see Fig. 6(b)).

Even though we focused on the dynamics of the splash curtain of the spheres, the model can be generalized to predict the splash curtain of projectiles with various geometries using the initial velocity of ejected splash and the rate of change of cavity volume.

This work was supported by the NSF under CBET-1336038.

## Appendix: Methods
### 1    Calculation of pressure difference force

The proposed mathematical model, describing the trajectory of the splash curtain requires knowledge of the pressure difference that exists across the splash, $\Delta P(t)$. It is hypothesized that this pressure difference is produced by air rushing into the expanding cavity created as the projectile penetrates downward into the fluid. Local low pressure in the cavity creates suction and pulls the splash radially inward, resulting in surface seal.

This pressure difference can be estimated by measuring the time rate of change of cavity volume, $dV_{cav}(t)/dt$. If the cavity is treated as a control volume, the rate of expansion (or collapse) can be thought of as the volumetric flow rate of air from the surrounding atmosphere into the cavity. If the area of the opening through which airflow is known, a mean air velocity, $U_{air}(t)$, can be estimated. With this mean air velocity, a pressure drop can be estimated using Bernoulli's principle.

The cavity was assumed to be axisymmetric about the $z$-axis (Fig. 3(a)). Integrating the cavity radius, $R(z,t)$, over the length of the cavity, $Z(t)$, at each time step yields the cavity volume history. An image processing routine was developed to determine the cavity profile from the captured image data and calculate the cavity volume, $V_{cav}(t)$.

As the sphere travels downward, it opens up and expanding the air cavity behind it. As the cavity expands, air is drawn into the cavity from the surrounding atmosphere. Neglecting compressibility effects, the time rate of change of cavity volume describes the volumetric flow rate of air into the cavity, given by

$$Q(t) = \frac{dV_{cav}(t)}{dt} \qquad (18)$$

where $Q(t)$ is the volumetric flow rate. In order to suppress the amplification of noise, which can often be a problem for numerical derivatives, the cavity volume data was lightly smoothed using robust locally weighted regression [30]. After the smoothing, a fourth order central difference scheme was used to compute the derivative [31].

If the area of the opening through which the air flows is known, a mean air velocity, $U_{air}(t)$, can be estimated. We already found the cavity profile for cavity volume determination, thus, we consider the opening radius, $R^*$, at each time instant, $t_i$, as the $R^*(t = t_i) = R(z = 0, t_i)$, and the cavity opening area is $\approx \pi R^{*2}$. Hence, the mean airflow velocity can be estimated by

$$U_{air}(t) \approx \frac{Q(t)}{\pi R^{*2}} \tag{19}$$

This provides an estimate of the mean air velocity flowing from the surrounding atmosphere, past the splash curtain, into the air cavity. We can use this condition to estimate a cavity pressure using Bernoulli's principle. Picking two arbitrary points along a streamline, one outside the cavity and the other inside the cavity, we can write

$$\left[\rho g z + \frac{1}{2}\rho U^2 + P\right]_{S_1} = \left[\rho g z + \frac{1}{2}\rho U^2 + P\right]_{S_2} \tag{20}$$

where $S_1$ is some position far away from the cavity and $S_2$ is a position just inside the cavity close to the rounded rim as seen in Fig. 3(a). Position $S_1$ was assumed to be at atmospheric pressure with no air motion. Position $S_2$ was assumed to be at cavity pressure, $P_{cav}(t)$, and have an airflow velocity of $U_{air}(t)$, given by Eq. **19**. Neglecting changes in elevation and assuming cavity conditions are constant throughout the cavity (not varying in space), the pressure difference across the splash can be estimated by

$$\Delta P(t) = P_{atm} - P_{cav}(t) = \frac{1}{2}\rho U_{air}^2(t) \tag{21}$$

where $\Delta P(t)$ is the pressure difference across the splash. This estimate assumes the air flowing into the cavity is incompressible (the density of air remains constant throughout the system, Mach number $\ll 1$), viscous affects are negligible, and we treat the cavity as a spatially uniform body. We also assume that $\Delta P(t)$ is spatially uniform within the cavity.

## 2  Splash curtain imaging

Finding the experimental trajectory of the splash curtain from a typical shadowgraph video similar to [13] is difficult and can impose high error in the rounded rim determination. Instead, we developed a variation of shadowgraph imaging by capturing the shadow of the splash curtain on a transparent high quality paper. To do so, we used a point light source to make the shadow of the splash on the transparent paper (using the point light source is crucial in order to prevent the creation of any penumbra and antumbra in the shadow, i.e. creation of sharp edge images), and focused the camera on the transparent paper instead of real splash curtain [32]. These images are then converted to binary images, enabled us to determine the rounded rim location of the splash curtain in a robust way: we obtained the splash curtain boundary profile and found the first inflection point from the bottom of it as the location of rounded rim in each frame [33].

We also found the rounded rim radius from the obtained curtain profile. This operation was performed for every frame of every case which resulted in a time history of rounded rim radius for each case. We averaged it over time for each individual drop and then calculated the mean and standard deviation for all the drops [33].